\documentclass[%
 reprint,
superscriptaddress,
 amsmath,amssymb,
 aps,
 prapplied,
floatfix,
]{revtex4-2}

\usepackage[colorlinks=true, allcolors=blue]{hyperref}
\usepackage[alsoload=synchem]{siunitx}
    \DeclareSIUnit \ev{eV}
    \DeclareSIUnit \Gauss{G}
\usepackage{braket}
\usepackage{graphicx}
\usepackage{dcolumn}
\usepackage{bm}
\usepackage{comment}

\begin{document}

\def \conceptFigure {
\begin{figure}
\includegraphics{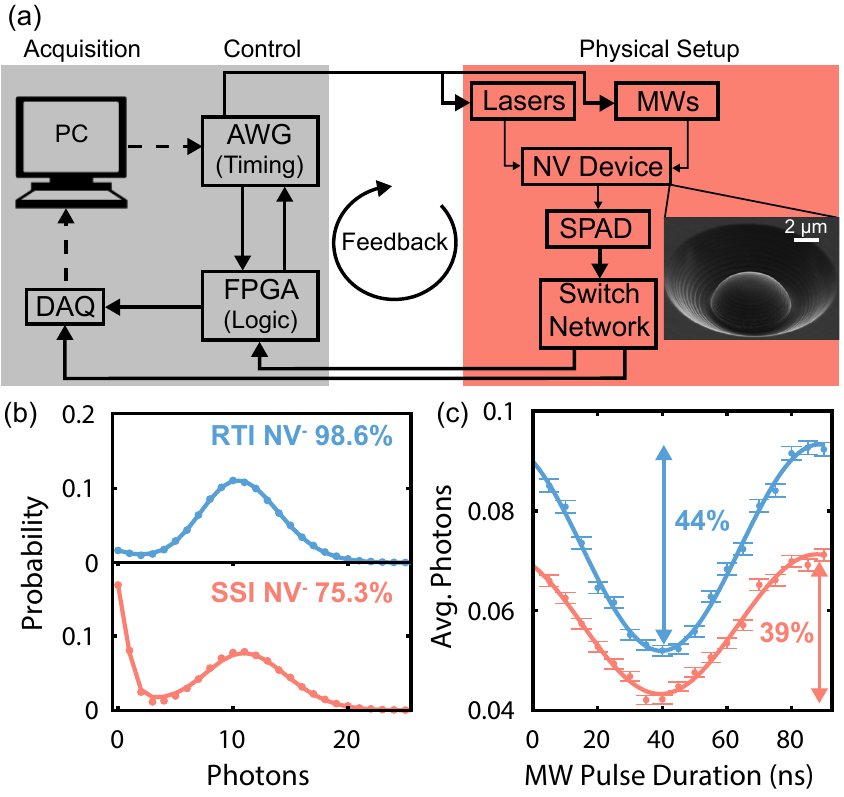}
\caption{\label{fig1} Real-Time Charge Initialization. (a) System overview for implementing real-time feedback on a nitrogen-vacancy (NV) center's charge state. Inset: scanning electron micrograph of a solid immersion lens fabricated around a single NV center. (DAQ: data acquisition, AWG: arbitrary waveform generator, FPGA: field programmable gate array, MWs: microwaves, SPAD: single-photon avalanche diode). (b) Charge-readout distributions demonstrating the difference in charge state initialization fidelity for the real-time (RTI, top panel) and steady state (SSI, bottom panel) initialization protocols. (c) Rabi nutations of a single NV center following RTI (top, blue curve and data points) and SSI (bottom, salmon curve and data points) demonstrating the increased signal and spin contrast (signified by the arrows). Curves are fits to a sinusoidal oscillation.}
\end{figure}
}

\def \realTimeModelFigure {
\begin{figure*}
\includegraphics{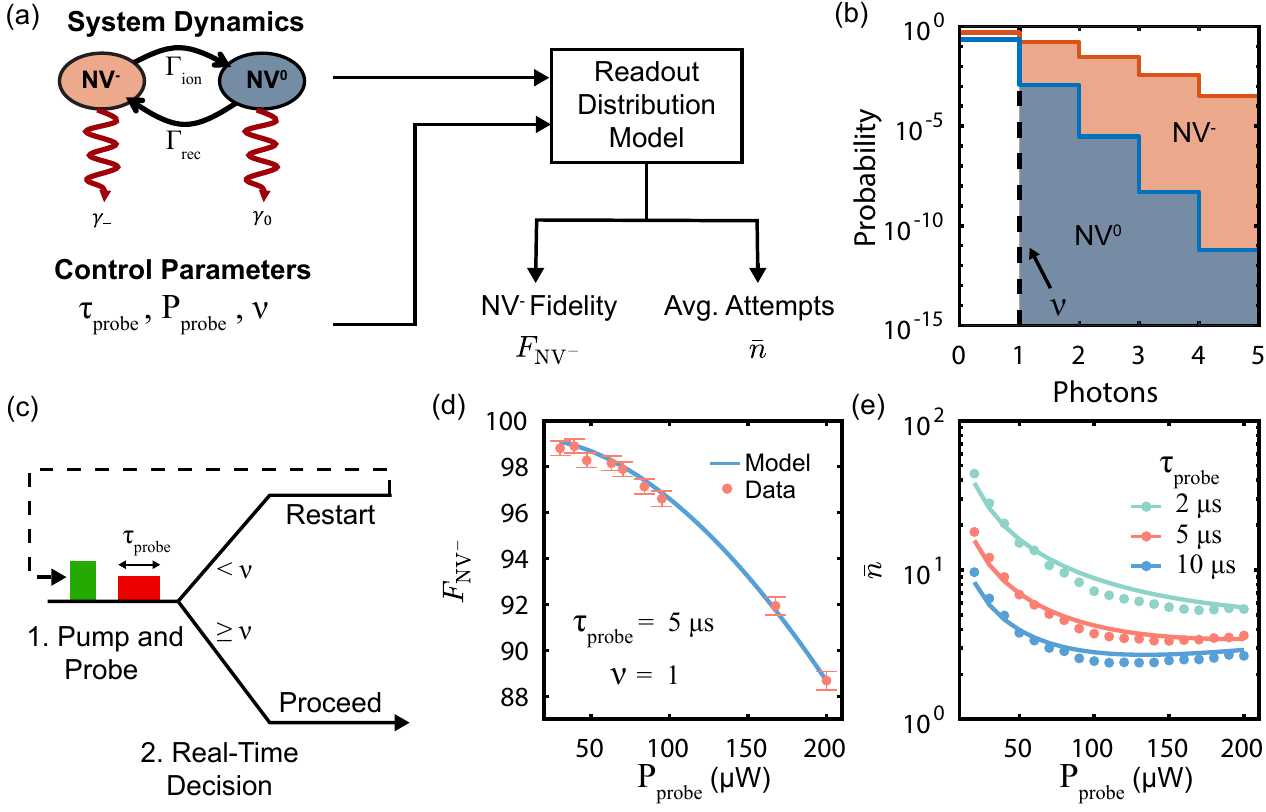}
\caption{\label{fig2} Modeling Real-Time Control. (a) Conceptual diagram of the model. The system dynamics model describes how the charge-dependent emission rates ($\gamma_-$, $\gamma_0$) and charge interconversion rates ($\Gamma_{\mathrm{ion}}$, $\Gamma_{\mathrm{rec}}$) depend on illumination power (P$_\mathrm{probe}$). Given a readout duration ($\tau_{\mathrm{probe}}$) and a threshold condition ($\nu_{\mathrm{threshold}}$), the readout distribution model determines the NV$^-$ fidelity and the average number of attempts required to reach the threshold. (b) Modeled photon distributions for the two charge states with $\tau_\mathrm{probe}=\SI{5}{\micro\second}$ and $P_\mathrm{probe}=\SI{100}{\micro\watt}$.
(c) Experimental timing diagram and decision tree for initializing the charge state. (d) Comparison between the modeled (line) and measured (markers) NV$^-$ fidelity as a function of probe power. (e) Comparison between the modeled (lines) and measured (markers) average attempts to reach the threshold ($\nu=1$) as a function of powers and probe duration. Errorbars in (e) are comparable to the marker size. }
\end{figure*}
}

\def \snrFidelityFigure {
\begin{figure}
\includegraphics{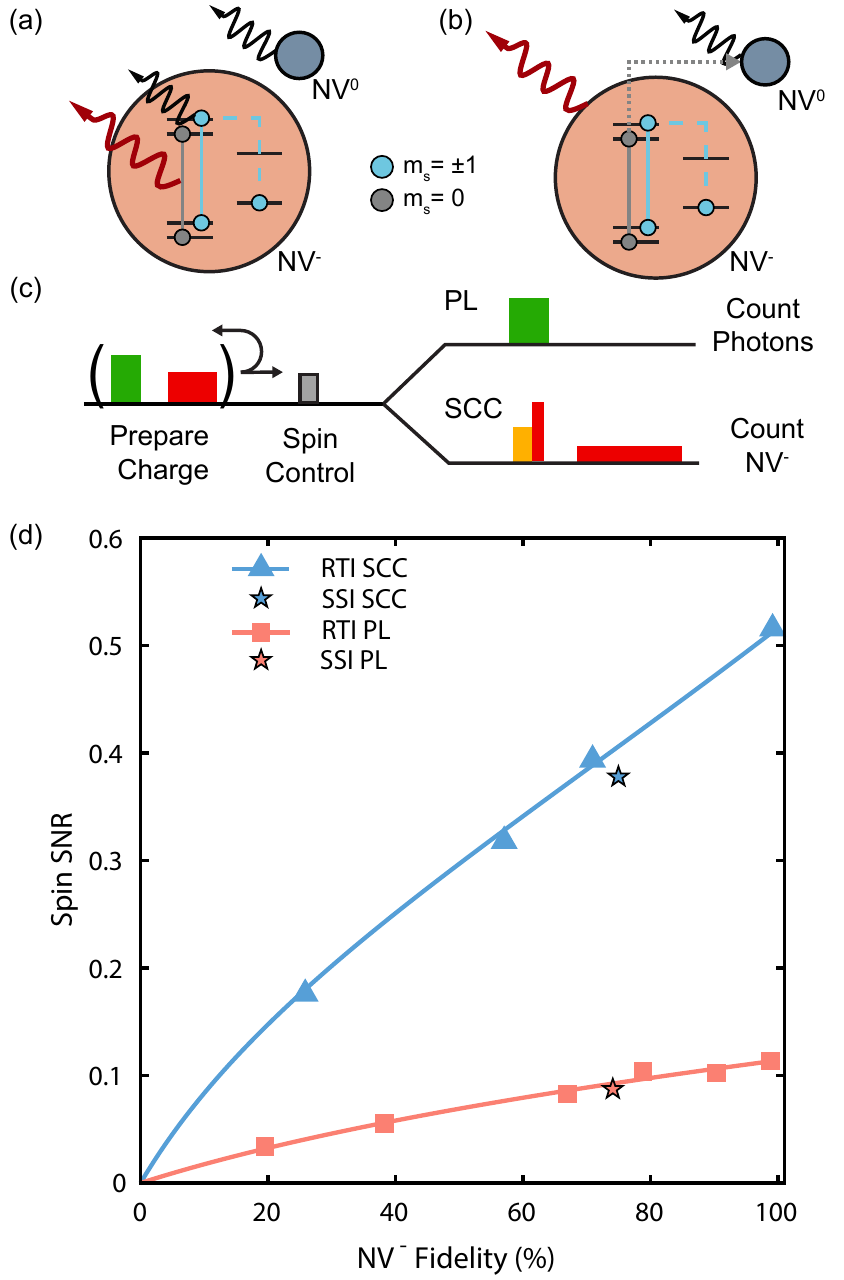}
\caption{\label{fig3} Charge Initialization Fidelity and Spin Readout (a,b) Conceptual diagram for traditional PL readout (a) and SCC (b). Red curves represent the desired signal and black curves contribute to background. (c) Timing diagram for measuring the spin signal-to-noise ratio (SNR) given different heralded charge fidelity. (d) Spin SNR as a function of NV$^-$ fidelity for different initialization and readout techniques.
The solid lines represent a fit of equation Eq.~\ref{eqn:expect_vs_fid} to the data, where the fit and data are converted to SNR using Eq.~\ref{eqn:ss_snr}.
Errorbars are comparable in size to the markers.}
\end{figure}
}

\def \speedupFigure {
\begin{figure*}
\includegraphics{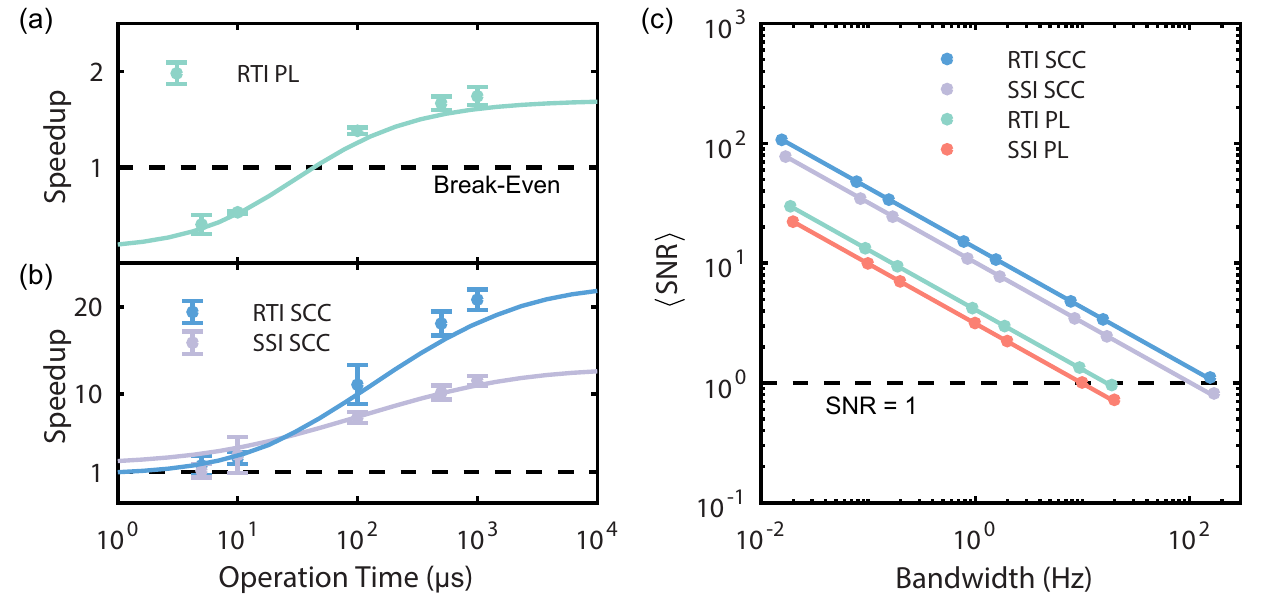}
\caption{\label{fig4} Readout Improvements Through Real-Time Initialization. Predicted (curves) and measured (data points) speedup  for PL spin readout with RTI (a) and SCC readout using SSI and RTI (b).
The dashed black line in panels (a) and (b) indicates the break-even condition in comparison to SSI and PL readout.
(c) Total, time-averaged SNR as a function of integration bandwidth for each protocol. Lines are fits to the inverse square root of the bandwidth. Errorbars in (c) are smaller than the markers.}
\end{figure*}
}

\preprint{AIP/123-QED}

\title{Real-Time Charge Initialization of Diamond Nitrogen-Vacancy Centers for Enhanced Spin Readout}

\def \ionRate {
\Gamma_\mathrm{Ion}
}

\def \recRate {
\Gamma_\mathrm{Rec}
}

\def \gammaMin {
\gamma_-
}

\def \gammaNeu {
\gamma_0}

\author{David A. Hopper}
\thanks{D.~A. Hopper and J.~D. Lauigan contributed equally to this work.}
\affiliation{
Quantum Engineering Laboratory, Department of Electrical and Systems Engineering, University of Pennsylvania, 200 S. 33rd St. Philadelphia, Pennsylvania, 19104, USA
}
\affiliation{
Department of Physics and Astronomy, University of Pennsylvania, 209 S. 33rd St. Philadelphia, Pennsylvania 19104, USA
}

\author{Joseph D. Lauigan}
\thanks{D.~A. Hopper and J.~D. Lauigan contributed equally to this work.}
\altaffiliation[Present address: ]{Quanergy Systems, 482 Mercury Dr, Sunnyvale, California 94085, USA}
\affiliation{
Quantum Engineering Laboratory, Department of Electrical and Systems Engineering, University of Pennsylvania, 200 S. 33rd St. Philadelphia, Pennsylvania, 19104, USA
}

\author{Tzu-Yung Huang}
\affiliation{
Quantum Engineering Laboratory, Department of Electrical and Systems Engineering, University of Pennsylvania, 200 S. 33rd St. Philadelphia, Pennsylvania, 19104, USA
}
\author{Lee C. Bassett}
\email[Corresponding author.  Email: ]{lbassett@seas.upenn.edu}
\affiliation{
Quantum Engineering Laboratory, Department of Electrical and Systems Engineering, University of Pennsylvania, 200 S. 33rd St. Philadelphia, Pennsylvania, 19104, USA
}

\date{\today}

\begin{abstract}
A common impediment to qubit performance is imperfect state initialization.
In the case of the diamond nitrogen-vacancy (NV) center, the initialization fidelity is limited by fluctuations in the defect's charge state during optical pumping.
Here, we use real-time control to deterministically initialize the NV center's charge state at room temperature.
We demonstrate a maximum charge initialization fidelity of 99.4$\pm$0.1\% and present a quantitative model of the initialization process that allows for systems-level optimization of the spin-readout signal-to-noise ratio.
Even accounting for the overhead associated with the initialization sequence, increasing the charge initialization fidelity from the steady-state value of 75\% near to unity allows for a factor-of-two speedup in experiments while maintaining the same signal-to-noise-ratio.
In combination with high-fidelity readout based on spin-to-charge conversion, real-time initialization enables a factor-of-20 speedup over traditional methods, resulting in an ac magnetic sensitivity of $\SI{1.3}{\nano\tesla\per\hertz^{1/2}}$ for our single NV-center spin.
The real-time control method is immediately beneficial for quantum sensing applications with NV centers as well as probing charge-dependent physics, and it will facilitate protocols for quantum feedback control over multi-qubit systems.

%
\end{abstract}

\maketitle

The accelerating pace of quantum technology is evident in the advancement of quantum sensors \cite{Degen2017} and the emergence of quantum networks \cite{Wehner2018}.
Critical to these developments have been solid-state spin qubits based on semiconductor defects, due to their optical interface \cite{Awschalom2018}, compatibility with integrated technologies \cite{Schroder:16}, and wide selection of host materials \cite{Atature2018}.
The most well-known example is the nitrogen-vacancy (NV) center in diamond \cite{Doherty2013,Hopper2018a}, which has enabled pivotal advances in quantum sensing \cite{Barry2016, Lovchinsky2017, Aslam2017a, Boss2017, Schmitt2017, Gross2017,Ariyaratne2018} and quantum information processing \cite{Hensen2015, Humphreys2018, Bradley2019}.

One limitation to the performance of NV-center qubits is imperfect initialization into the oft-desired negative charge state (NV$^-$).
Optical pumping with 532~nm light produces a steady-state statistical charge distribution; typically the probability to prepare the NV$^-$ state is around 75\% \cite{Waldherr2011a, Aslam2013}, although it can be much lower for defects close to surfaces \cite{Bluvstein2019}.
This probabilistic steady-state initialization (SSI) hampers spin readout by decreasing contrast and increasing readout noise \cite{Hopper2018a}, and it limits the fidelity of quantum gate operations of coupled spin systems utilizing the NV center as an ancilla \cite{Neumann2010, Waldherr2011a}.
Existing techniques to improve the charge initialization fidelity include doping electrically \cite{Doi2014} or chemically \cite{Doi2016}, and multi-color optical pumping \cite{Hopper2016}.
In addition, many experiments utilize post selection to filter out the noise \cite{Waldherr2011, Waldherr2012, Waldherr2014, Shields2015, Hopper2016, Bluvstein2019}.
These techniques either impose strict constraints on materials and device design or require elongated experimental runtime.
At cryogenic temperatures, deterministic initialization protocols based on real-time feedback have been essential for entanglement generation and quantum error correction using NV centers due to their long measurement times \cite{Bernien2013, Cramer2016, Humphreys2018}, however these techniques have not been adapted for quantum sensing applications where the duration of each measurement cycle drastically affects the overall sensitivity.

Here, we use real-time feedback to control an NV center's charge-state initialization fidelity at room temperature, and we demonstrate improved spin readout efficiency and sensitivity [Fig.~\ref{fig1}].
A model for the stochastic initialization procedure allows for the selection of near-unity initialization fidelity into either charge state, or an arbitrary intermediate charge distribution.
We measure the influence of charge fidelity on the spin readout signal-to-noise ratio (SNR) for two readout techniques, traditional photoluminescence (PL) and spin-to-charge conversion (SCC).
Our comprehensive model allows for the optimization of initialization and readout parameters for quantum control experiments of arbitrary durations.
The real-time initialization (RTI) protocol improves the spin readout efficiency and reduces the time required for experiments; in combination with SCC readout, we demonstrate a factor-of-20 speedup as compared to traditional methods.

A schematic of the experiment is shown in Fig.~\ref{fig1}(a).
The traditional portion of the setup consists of the lasers, microwave sources, diamond device, and photon-counting electronics.
The sample is an electronics grade type-IIa synthetic diamond (Element Six) which has been irradiated with \SI{2}{\mega\ev} electrons ($10^{14}$ \si{\centi\meter}$^{-2}$) and annealed at $800^\circ$C for 1 hour in forming gas.
A solid immersion lens aligned to a single NV center was fabricated using focused-ion-beam milling to increase collection efficiency \cite{Hopper2016}.
Imaging and optical control is performed with a home-built room-temperature scanning confocal microscope with three excitation sources.
A continuous-wave \SI{532}{\nano\meter} laser (Gem 532, Laser Quantum), referred to as ``green,'' is gated by an acousto-optic modulator (AOM) in a double-pass configuration; it is used for optical pumping and traditional PL readout.
An amplitude modulated \SI{635}{\nano\meter} laser diode (MLD 06-01 638, Cobolt), referred to as ``red,'' is used for charge readout and SCC.
A continuous-wave \SI{592}{\nano\meter} laser (VFL-592, MPB Communications, Inc.), referred to as ``orange,'' is gated with an AOM and is used for SCC.
A \SI{115}{\Gauss} magnetic field is aligned along the NV axis to distinguish the $m_s=\pm1$ states.
A lithographically-defined loop-antenna surrounding the solid immersion lens is driven by an amplified (ZHL-16W-43-S+, Mini Minicircuits), amplitude modulated (ZASWA-2-50DR, Mini Circuits), continuous-wave signal generator (SG384, Stanford Research Systems), which allows for ground-state spin control.

\conceptFigure

The NV charge state is determined to a high accuracy by utilizing a wavelength that excites the NV$^-$ zero phonon line of \SI{637}{\nano\meter} but not NV$^0$ zero phonon line of \SI{575}{\nano\meter}.
Example histograms of photon counts arising from 75,000 charge readouts are shown in Figure.~\ref{fig1}(b) for both the steady-state NV$^-$ population of $75.3\pm0.4\%$ and a higher fidelity initial population of $98.6\pm0.2\%$.
The benefit of this elevated initialization fidelity can be seen in the ground-state Rabi nutations in Fig.~\ref{fig1}(c), where the higher fidelity charge state exhibits higher brightness and contrast.

We implement real-time control by linking our timing electronics, which consist of an arbitrary waveform generator (AWG, AWG520 Tektronix) and data acquisition (DAQ, National Instruments) system, with the fast digital logic of a field programmable gate array (FPGA, Virtex-7 Xilinx); refer to Figure~\ref{fig1}(a) for the full system overview.
In the initialization control loop, the AWG outputs a sequence consisting of a green pump and red charge probe in an repeating loop; when the FPGA detects that a preset photon detection threshold has been reached duing the charge probe, it sends an event signal to advance the AWG out of its loop and continue with the other predefined measurements.
The time it takes from detection of the final photon to the halting of the initialization procedure is $\tau_\mathrm{delay}=\SI{550}{\nano\second}$, which consists of the detector delay (\SI{30}{\nano\second}), the AWG delay (\SI{500}{\nano\second}), and the red laser delay (\SI{20}{\nano\second}).

\realTimeModelFigure

We model the charge probe process using a photon  distribution model accounting for transitions between NV$^-$ and the neutral (NV$^0$) charge state \cite{Shields2015, DAnjou2016, Hacquebard2018}.
The model assumes that the charge dynamics of the NV center can be reduced to a two-state system with emission rates $\gammaMin{}$ and $\gammaNeu{}$, and charge transition rates for ionization (negative to neutral, $\ionRate$) and recombination (neutral to negative, $\recRate{}$); see Fig.~\ref{fig2}(a).
We determine these rates as a function of power by measuring the photon distributions during a time bin that allows for about one ionization event to occur and fitting to the model \cite{Supplemental}.
Since the charge readout powers used in this work are below the saturation regime, the emission rates scale linearly with laser power while the ionization and recombination rates scale quadratically with power  \cite{Waldherr2011a, Aslam2013}.

The control parameters governing the charge probe process are the laser power (P$_\mathrm{probe}$), maximum duration ($\tau_\mathrm{probe}$), and the photon threshold  ($\nu$) that breaks the initialization loop [Fig.~\ref{fig2}(a)].
Given these three parameters, the model provides the expected photon distributions for the negative or neutral charge state configurations,
\begin{equation}
    p(n | s),
\end{equation}
where $n$ is the number of photons detected during $\tau_\mathrm{probe}$ and $s=-$ or $0$ signifies the initial charge state; see Fig.~\ref{fig2}(b) for an example.

The distributions allow us to calculate two critical metrics for RTI: the NV$^-$ charge fidelity ($F_{\mathrm{NV}^-}$) and the average attempts ($\bar{n}$) required for successful initialization.
The initialization fidelity is governed by two terms,
\begin{equation}
    F_{\mathrm{NV}^-} = (1-\epsilon_T)(1-\epsilon_{D}),
\end{equation}
where $\epsilon_T$ is the threshold error and $\epsilon_D$ is the delay error.
The threshold error is the probability that NV$^0$ leads to a threshold reaching event and is given by
\begin{equation}
   \epsilon_\mathrm{T} =  \frac{\sum_{n \geq \nu}(1-P_-)p(n | 0)}{\sum_{n \geq \nu}P_-p(n | -) + (1-P_-)p(n | 0)},
\end{equation}
where $P_-$ is the probability that the NV center was initially in NV$^-$ prior to the charge probe.
The delay error is the probability that an ionization event occurred during the electronic delay time and is given by
\begin{equation}
    \epsilon_D = 1 - e^{-\tau_\mathrm{delay}\ionRate{}}.
\end{equation}
%
The average attempts to initialize is given by
\begin{equation}
\bar{n} = \left(\sum_{n \geq \nu}P_-p(n | -) + (1-P_-)p(n | 0) \right)^{-1}.
\end{equation}
As an ensemble average, $\bar{n}$ takes continuous values.

Fig. ~\ref{fig2}(c) outlines the experimental decision tree in the real-time initialization procedure.
A charge pump-and-probe sequence is repeatedly played out by the AWG until the FPGA detects a threshold-reaching event.
The green pump pulse is set to \SI{500}{\micro\watt} and \SI{500}{\nano\second} to quickly repump the charge without incurring significant overhead; we vary $P_\mathrm{probe}$ and $\tau_\mathrm{probe}$ to optimize the performance.
In order to verify our model, we measure $F_{\mathrm{NV}^-}$ and $\bar{n}$ as a function of $P_\mathrm{probe}$ as shown in Figs.~\ref{fig2}(d) and (e).
We extract $F_{\mathrm{NV}^-}$ by performing a subsequent charge measurement and fitting to the photon distribution model, and determine $\bar{n}$ from the time it takes to record 10$^5$ threshold reaching events.

The measurements of $F_{\mathrm{NV}^-}$ are generally consistent with our model.
We attribute the minor discrepancy between the measured values of $\bar{n}$ and the model predictions to minor variations in the steady-state charge population imposed by the control sequence.
The model assumes an initial NV$^-$ population of $P_-=75\%$ prior to each probe, however
We neglect this higher-order effect since it has the beneficial affect of decreasing $\bar{n}$ for the control parameters we employ.

The relative contribution of the two error sources in the charge initialization fidelity depend on the power.
At very low powers, $\epsilon_T$ is dominant and $F_{\mathrm{NV}^-}$ is limited by the signal-to-background ratio of the charge readout process.
At higher powers, $\epsilon_D$ is dominant due to the quadratic scaling of the ionization rate with power.
Therefore, when designing an experiment utilizing RTI, it is crucial to minimize the control delay time.
For a threshold of 1 photon, the maximum achievable fidelity is $98.6\pm0.2\%$ at low powers [Fig.~\ref{fig2}(d)].

To verify that RTI preserves the ground state spin properties, we measured the coherence times for a Ramsey ($T_2^*)$ and Hahn echo ($T_2$) measurement, as well as the spin relaxation time ($T_1$) \cite{Supplemental}.
We observe a $\sim16\%$ increase in $T_2^*$ when utilizing RTI, which could be due to ionization of nearby substitutional nitrogen donors \cite{Doherty2016}, but we detect no statistically significant difference in $T_2$ or $T_1$.

We now consider the effect of the initial $F_{\mathrm{NV}^-}$ on the spin readout SNR.
Generally, the observable for a spin measurement of an NV center follows the form
\begin{align}
    \braket{S_i} = \braket{\tilde{S}_i}F_{\mathrm{NV}^-} + \braket{\epsilon}(1-F_{\mathrm{NV}^-}),
    \label{eqn:expect_vs_fid}
\end{align}
where $\braket{S_i}$ is ensemble-averaged value of the observable $S$ for the spin state $i$, $\braket{\tilde{S}_i}$ is the expectation value of the observable for spin state $i$ given an initial NV$^-$ state, and $\braket{\epsilon}$ is an error in the observable which is due to the NV center residing in NV$^0$ during the readout.
The single-shot SNR for spin readout is then given by
\begin{align}
    \mathrm{SNR} = \frac{\left|\braket{S_0} - \braket{S_1}\right|}{\sqrt{\sigma_0^2 + \sigma_1^2}},
    \label{eqn:ss_snr}
\end{align}
where $\sigma_i$ is the standard deviation associated with $\braket{S_i}$ \cite{Hopper2018a}.

To make quantitative comparisons between readout techniques, the physical observable and its accompanying statistical model must be incorporated into equation (\ref{eqn:ss_snr}).
For PL readout [Fig.~\ref{fig3}(a)], the signal is the average number of detected photons during the first \SI{250}{\nano\second} of \SI{532}{\nano\meter} illumination and thus obeys Poissonian statistics.
For SCC readout [Fig.~\ref{fig3}(b)], the signal is the probability of detecting NV$^-$ following the conversion, and it obeys Binomial statistics.

Figure~\ref{fig3}(c) details the measurement timing diagram that allows for the characterization of spin SNR as a function of $F_{\mathrm{NV}^-}$.
Following initialization with an arbitrary $F_{\mathrm{NV}^-}$, the spin state is either left in the polarized $m_s=0$ state, or flipped to the $m_s=-1$ state with a \SI{40}{\nano\second} microwave $\pi$-pulse.
We estimate the value of $\braket{S_i}$ from repeated measurements using both traditional and SCC readout techniques.
We also measure the spin SNR for the traditional SSI consisting of \SI{2}{\micro\second} of \SI{532}{\nano\meter} illumination.
We separately optimize PL and SCC readout parameters to ensure a fair comparison between the techniques \cite{Supplemental}.
The raw data are fit using equation~(\ref{eqn:expect_vs_fid}), from which we empirically determine $\braket{\tilde{S}_i}$ and $\braket{\epsilon}$.
Figure~\ref{fig3}(d) depicts the results of this measurement for both readout protocols, with the SNR calculated using equation~(\ref{eqn:ss_snr}) for both the data (symbols) and fits (curves).

Interestingly, the spin SNR following RTI for both SCC and PL readout, when controlling for NV$^-$ fidelity, is $\sim7\%$ higher than for SSI.
This is attributed to improved optical spin polarization in the real-time protocol, since the red laser induces negligible recombination; this is consistent with previous observations \cite{Chen2015}.
The initial spin purity, estimated from measurements of the excited-state lifetime, is approximately 91\% and 94\% for the steady state and real-time protocols, respectively \cite{Supplemental}.

\snrFidelityFigure

\speedupFigure

By combining the RTI model with the spin SNR as a function of $F_{\mathrm{NV}^-}$, we can optimize the signal acquisition for a given experiment.
To achieve this, we define the readout efficiency,
\begin{align}
    \xi = \frac{\mathrm{SNR}}{\sqrt{\tau_I + \tau_O + \tau_R}},
    \label{eqn:snr_root_time}
\end{align}
where $\tau_I$ is the initialization time, $\tau_O$ is the spin operation time, and $\tau_R$ is the spin readout time.
This figure of merit is related to the sensitivity, and encompasses the single-shot SNR, the spin operation duration, and the associated initialization and readout overheads \cite{Hopper2018a}.
The total SNR after multiple measurement cycles with a total integration time, $T$, is simply given by  $\braket{\mathrm{SNR}}=\xi\sqrt{T}$.
We assume the operation time is fixed by the desired sensing or computation protocol.
We have previously considered the optimization of the readout duration, power, and threshold for SCC, and we include those procedures when necessary \cite{Hopper2018, Hopper2018a}.

Real-time control allows for additional design flexibility in an experiment, as longer time spent initializing results in a higher spin readout SNR yet fewer total averages.
Equation~(\ref{eqn:snr_root_time}) quantitatively captures the trade-off between these two quantities.
The initialization time is given by
\begin{align}
    \tau_I = (\tau_{\mathrm{pump}} + \tau_{\mathrm{overhead}} + \tau_{\mathrm{probe}})\bar{n},
\end{align}
where $\tau_{\mathrm{pump}}=\SI{0.5}{\micro\second}$ is the duration of the \SI{532}{\nano\meter} charge reset pump and $\tau_{\mathrm{overhead}}=\SI{1.5}{\micro\second}$ is the overhead in the initialization sequence omprised of the green AOM delay, singlet decay time, and $\tau_\mathrm{delay}$.
Note that $\tau_I$ is an average quantity since equation~(\ref{eqn:snr_root_time}) is assumed to be an ensemble average over many trials.

With a model describing the readout efficiency, we can numerically optimize equation~(\ref{eqn:snr_root_time}) to determine the protocol parameters that maximize the readout efficiency for a given operation time.
To assess the results in context of typical NV-center experiments, we compute and measure the baseline readout efficiency, $\xi_\mathrm{baseline}$, corresponding to steady state initialization and traditional PL readout for different operation times.
We then define the speedup as the reduction in integration time required to achieve a fixed SNR when comparing a new technique to the baseline,
\begin{align}
    \mathrm{Speedup} = \left(\frac{\xi}{\xi_{\mathrm{baseline}}}\right)^2.
\end{align}
A speedup of unity defines the break-even time; the operation time at which it is equally efficient to use the enhanced technique over the baseline protocol.

Figure~\ref{fig4} presents the results of this optimization for four different scenarios: SSI with PL readout, RTI with PL readout, SSI with SCC readout, and RTI with SCC readout. 
The predicted and measured speedup curves for PL and SCC readout are shown in Figs.~\ref{fig4}(a) and (b), respectively.
For PL readout, we observe a break-even time for using the RTI of $\sim\SI{70}{\micro\second}$,
and a maximum speedup of 1.74$\pm0.09$ for an operation time of \SI{1}{\milli\second}.
Interestingly, we find that our full model always results in a choice of measurement parameters that make SCC more efficient than PL readout.
RTI offers a further boost for operation times over \SI{30}{\micro\second}, with a maximum observed speedup of 20.8$\pm1.2$ for $\tau_O=\SI{1}{\milli\second}$.
The measurements slightly out-perform the model at long operation times; this can likely be explained by uncertainty in the calibration measurements that underlie the model curves.


Figure~\ref{fig4}(c) shows the total SNR as a function of integration bandwidths for each of the four techniques.
Here, we have fixed the operation time to be \SI{500}{\micro\second}.
In each case, the total SNR scales with the inverse square root of bandwidth as expected.
Of note is the integration bandwidth for which each technique achieves $\braket{\text{SNR}}=1$, which represents the maximum frequency of environmental dynamics that can be resolved above the noise.
The RTI protocol coupled with SCC readout offers the best performance for this operation time.
In addition, Fig.~\ref{fig4}(c) confirms that the optical pulse sequences required for RTI and SCC do not introduce any appreciable noise in the bandwidth we consider.

NV-center quantum sensors stand to gain significant sensitivity improvements from using RTI protocols of the charge state.
The largest speedup is realized for long operation times that approach \SI{1}{\milli\second}, which coincide with the typical requirements for spin relaxometry \cite{Pelliccione2014, Ariyaratne2018} as well as dynamical decoupling sequences \cite{Taminiau2012, Kolkowitz2012}.
The single NV center studied here exhibits a Hahn-echo $T_2=\SI{800}{\micro\second}$.
Explicitly accounting for the RTI and SCC overhead, our observed $\xi$ corresponds to an ac magnetic sensitivity of $\SI[]{1.3}{\nano\tesla \per\hertz^{1/2}}$ \cite{Supplemental}.
Other readout techniques used in quantum sensors, such as the nuclear-assisted method \cite{Haberle2017, Aslam2017a}, would also see similar signal-acquisition improvements due to RTI.
In many situations, the gains are likely to be even larger than we have demonstrated, since NV centers located in nanodiamonds or close to surfaces typically exhibit lower steady-state charge populations than those in bulk diamond \cite{Hopper2018,Bluvstein2019}.
For example, using our model and assuming a 25\% NV$^-$ steady-state population, RTI would enable a factor-of-6 speedup for PL readout and a factor-of-75 speedup for SCC readout with an operation time of \SI{500}{\micro\second}.

While we have focused on applications that require strict consideration of the overhead from initialization and readout, the techniques are directly applicable to situations in which initialization fidelity is prioritized over total measurement time.
For example, the initialization error can be reduced by a further factor of 2 by increasing the threshold to 2 photons, and the delay error can be reduced by decreasing $P_\mathrm{probe}$, which leads to  $F_\mathrm{NV^-}=99.4\pm0.1\%$ with $\tau_I=\SI{7}{\milli\second}$ \cite{Supplemental}.
Such control over the charge state could facilitate precise measurements of the local electrostatic environment \cite{Mittiga2018, Bluvstein2019}, aid in the quantification of photon collection efficiency for photonic devices \cite{Huang2019}, and improve the single-shot SNR for infrequent SCC measurements \cite{Jaskula2019}.
In addition, the fidelity associated with initializing, controlling, and measuring coupled nuclear spins \cite{Neumann2010,Cramer2016,Bradley2019} is intricately tied to the NV center's charge and spin purity and thus could be improved with RTI.

In conclusion, we demonstrated an efficient method for initializing the charge state of an NV center in real-time and assessed how this can be used to improve the efficiency of spin readout.
Real-time control could be applied to other aspects of the NV center, such as projective initialization of nuclear spins \cite{Liu2017} and increasing the spin state initialization fidelity through time-gating.
In addition, this advanced control can be applied to other emerging solid-state spin defects, especially those which may have a high fidelity readout mechanism but a less-than-ideal spin or charge pumping transition.

\section*{Acknowledgements}
This work was supported by the National Science Foundation under awards ECCS-1553511 (DAH, JDL, and LCB) and ECCS-1842655 (TYH and LCB). The authors thank H. S. Knowles and Z. Zhang for useful discussions and D. Bluvstein, S. A. Breitweiser, and R. J. Patel for valuable comments on the manuscript.

\bibliography{charge-init}

\end{document}



\title{Supplementary Information for ``Real-Time Charge Initialization of Diamond Nitrogen-Vacancy Centers for Enhanced Spin Readout''} 

\def \ionRate {
\Gamma_\mathrm{Ion}
}

\def \recRate {
\Gamma_\mathrm{Rec}
}

\def \gammaMin {
\gamma_-
}

\def \gammaNeu {
\gamma_0}

\renewcommand{\figurename}{SFIG.}



\author{David A. Hopper}
\altaffiliation{These authors contributed equally.}
\affiliation{ 
Quantum Engineering Laboratory, Department of Electrical and Systems Engineering, University of Pennsylvania, 200 S. 33rd St. Philadelphia, Pennsylvania, 19104, USA
}
\affiliation{
Department of Physics and Astronomy, University of Pennsylvania, 209 S. 33rd St. Philadelphia, Pennsylvania 19104, USA
}

\author{Joseph D. Lauigan}
\altaffiliation{These authors contributed equally.}
\affiliation{ 
Quantum Engineering Laboratory, Department of Electrical and Systems Engineering, University of Pennsylvania, 200 S. 33rd St. Philadelphia, Pennsylvania, 19104, USA
}
\author{Tzu-Yung Huang}
\affiliation{
Quantum Engineering Laboratory, Department of Electrical and Systems Engineering, University of Pennsylvania, 200 S. 33rd St. Philadelphia, Pennsylvania, 19104, USA
}
\author{Lee C. Bassett}
\altaffiliation{
Correspondence: lbassett@seas.upenn.edu
}
\affiliation{ 
Quantum Engineering Laboratory, Department of Electrical and Systems Engineering, University of Pennsylvania, 200 S. 33rd St. Philadelphia, Pennsylvania, 19104, USA
}


\date{\today}


\pacs{}

\maketitle 


\section{Real-Time Control Hardware}
Critical to this experiment is the ability for our timing electronics (AWG520, Tektronix) to be able to exit an infinite loop asynchronously and move on to a different timing sequence conditioned on an external signal.
The Xilinx Artix-7 FPGA is physically interfaced with a Digital ARTY development board and perform the real-time counting and logic necessary.
This platform exposes the I/O headers to communicate with the digital logic and upload the board firmware.
We define the digital logic in Verilog and use the Vivado IDE to build and communicate the instructions to the board.
All of our counting, both for initialization and readout, is handled by the FPGA. 
Given the type of counting, the FPGA either triggers an event to break out of the initialization loop or forwards the number of counts and the trial number to the DAQ via 12 digital lines after a readout has finished.
The core functionality of the real-time control can be explained by three modules: control, SPAD counter, and trial counter. The control module receives information via digital lines from the AWG that governs when the FPGA should be counting, what threshold to check against, resetting registers, and whether an event should be triggered (for initialization) or if the threshold should be ignored (for readout). 
The SPAD counter is a 6-bit counter that records the number of rising edges coming from the SPAD and handles resetting the number of counts for a given task.
The trial counter is similar to the SPAD counter, but it counts rising edges originating from the AWG.
The two counters are output as a 12-bit register that the DAQ samples following the end of an experiment.
This dual-counter strategy allows us to run multiple trials of a given experiment in an interleaved fashion.
An example Verilog file that was used in our experiment has been posted publicly on GitHub \cite{github}.

\section{Charge Readout Model}
We use the simplified photon distribution function originally proposed by \citet{Shields2015}, which models the observed photon distribution produced by an NV center undergoing charge readout.
This model includes the effects of ionization and recombination during the readout, allowing for accurate determination of the initial NV$^-$ population.
Given the NV is initially in NV$^-$, the photon probability distribution depends on whether there were an odd or even number of charge transitions, and is given by
\begin{multline}
    p(n|-, \mathrm{odd}) = \int_0^{t_R} d\tau e^{(\recRate - \ionRate)\tau -\recRate{t_R}}
    \times \ionRate{} \times\mathrm{BesselI}\left(0, 2\sqrt{\ionRate\recRate\tau(t_R-\tau)}\right)\times \\
    \mathrm{PoissPDF}(\gammaMin\tau + \gammaNeu(t_R - \tau), n)
    \label{eqn:distr_even}
\end{multline}
\begin{multline}
    p(n|-, \mathrm{even}) = \int_0^{t_R} d\tau e^{(\recRate - \ionRate)\tau -\recRate{t_R}}
    \times \sqrt{\frac{\ionRate\recRate\tau}{t_R-\tau}}\times \\ \mathrm{BesselI}\left(1, 2\sqrt{\ionRate\recRate\tau(t_R-\tau)}\right)\times 
    \mathrm{PoissPDF}(\gammaMin\tau + \gammaNeu(t_R - \tau), n) \\
    + e^{-\ionRate t_R}\mathrm{PoissPDF}(\gammaMin t_R, n),
    \label{eqn:distr_odd}
\end{multline}
where $\gammaMin{}$ ($\gammaNeu{}$) is the photon detection rate due to NV$^-$ (NV$^0$), $\ionRate{}$ ($\recRate{}$) is the ionization (recombination) rate, $t_R$ is the total readout duration, and BesselI is a modified Bessel function of the first kind. 
The full photon distribution given starting in NV$^-$ is 
\begin{align}
    p(n|-) = \frac{1}{2}(p(n|-, \mathrm{odd}) + p(n|-, \mathrm{even})).
    \label{eqn:distribution_model}
\end{align}
To get the equivalent expression for NV$^0$, simply swap the ionization and recombination rates as well as the photon rates ($\ionRate \leftrightarrow \recRate$ and $\gammaMin{} \leftrightarrow \gammaNeu{}$) in Eqs.~\ref{eqn:distr_even} and \ref{eqn:distr_odd}.
A general mixture of the two charge states is given by
\begin{align}
    p(n) = P_-p(n|-) + (1-P_-)p(n|0),
    \label{eqn:general_charge_distribution}
\end{align}
where $P_-$ is the initial population of NV$^-$ prior to the start of readout.
In this work, we evaluate Eq.~\ref{eqn:general_charge_distribution} numerically in MATLAB and either fit to measured distributions or simulate the expected photon distribution for a given choice of rates and readout duration.

\section{Red Charge Dynamics Calibration}

\begin{figure}
    \centering
    \includegraphics{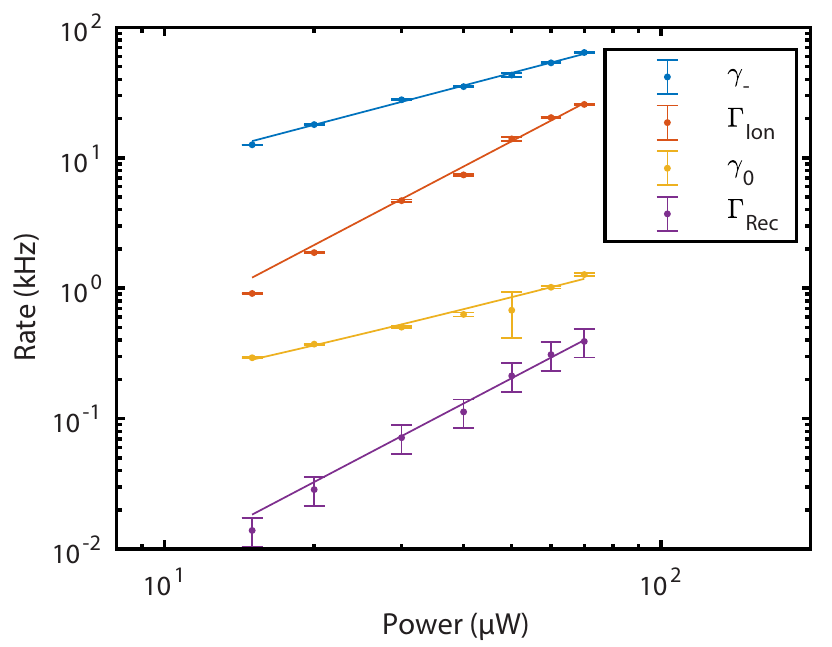}
    \caption{Red Charge Dynamics Calibration. The four measured rates and their associated power-dependence fits are presented.}
    \label{sSFig:charge_dynamics_data}
\end{figure}

To calibrate the red charge dynamics, we fit measured photon distributions to the model previously described for different red powers.
The readout duration is chosen such that $t_R > \ionRate{}^{-1}$, allowing for the effects of ionization to be observed.
The readout duration does not need to be longer than the recombination rate, as it is is given by
\begin{equation}
    \Gamma_{\mathrm{rec}} =\frac{P_{-}}{1-P_-}\Gamma_\mathrm{ion}.
\end{equation}
Due to the small recombination rate under red illumination, the steady state NV$^-$ population is $P_- = 1.15\pm0.07\%$.
This complicates measuring the charge dynamics as it is difficult to observe NV$^-$ fluorescence during continuous red illumination.
To overcome this, we implement a pump-probe scheme to determine the red charge dynamics.
A \SI{10}{\micro\second} green charge pump is followed by a variable power and duration red charge probe during which the number of photons are counted.
This increases the average initial NV$^-$ population to $\approx75\%$.
The results of the charge dynamics calibration are presented in SFigure~\ref{sSFig:charge_dynamics_data}.
We fit all of the rates to their unsaturated form \cite{Aslam2013} due to the maximum power being well below saturation. 
The power-dependencies for the rates are as follows
\begin{align}
    \gamma_- = C_-P \\
    \gamma_0 = C_0P + D \\
    \Gamma_\mathrm{Ion} = C_\mathrm{Ion}P^2 \\
    \Gamma_\mathrm{Rec} = C_\mathrm{Rec}P^2,
\end{align}
where $C_s$ is the rate scaling term for the $s$ process and $D$ is the dark count rate.
The fit parameters are presented in Table~\ref{table:charge_fit}.
The agreement of the data with the fit model further supports the use of the unsaturated rate dependencies.

\begin{table}
\centering
\begin{tabular}{| c | c |}
\hline
Parameter & Value \\
\hline \hline
     $C_-$ & $0.895\pm0.027$ \si{\kilo\hertz} \si{\micro\watt}$^{-1}$ \\
     $C_0$ &  $0.0163\pm0.0023$  \si{\kilo\hertz} \si{\micro\watt}$^{-1}$\\
     $D$ &  $0.039\pm0.067$  \si{\kilo\hertz}\\
     $C_\mathrm{Ion}$ & $5.36\pm0.27$ \si{\hertz} \si{\micro\watt}$^{-2}$ \\
     $C_\mathrm{Rec}$ & $0.082\pm0.0041$ \si{\hertz} \si{\micro\watt}$^{-2}$ \\ 
     \hline
\end{tabular}
\caption{Fit results for the charge dynamics calibration.}
\label{table:charge_fit}
\end{table}

\section{Spin Properties Following Real-Time Initialization}
\begin{figure}
\includegraphics[scale=0.95]{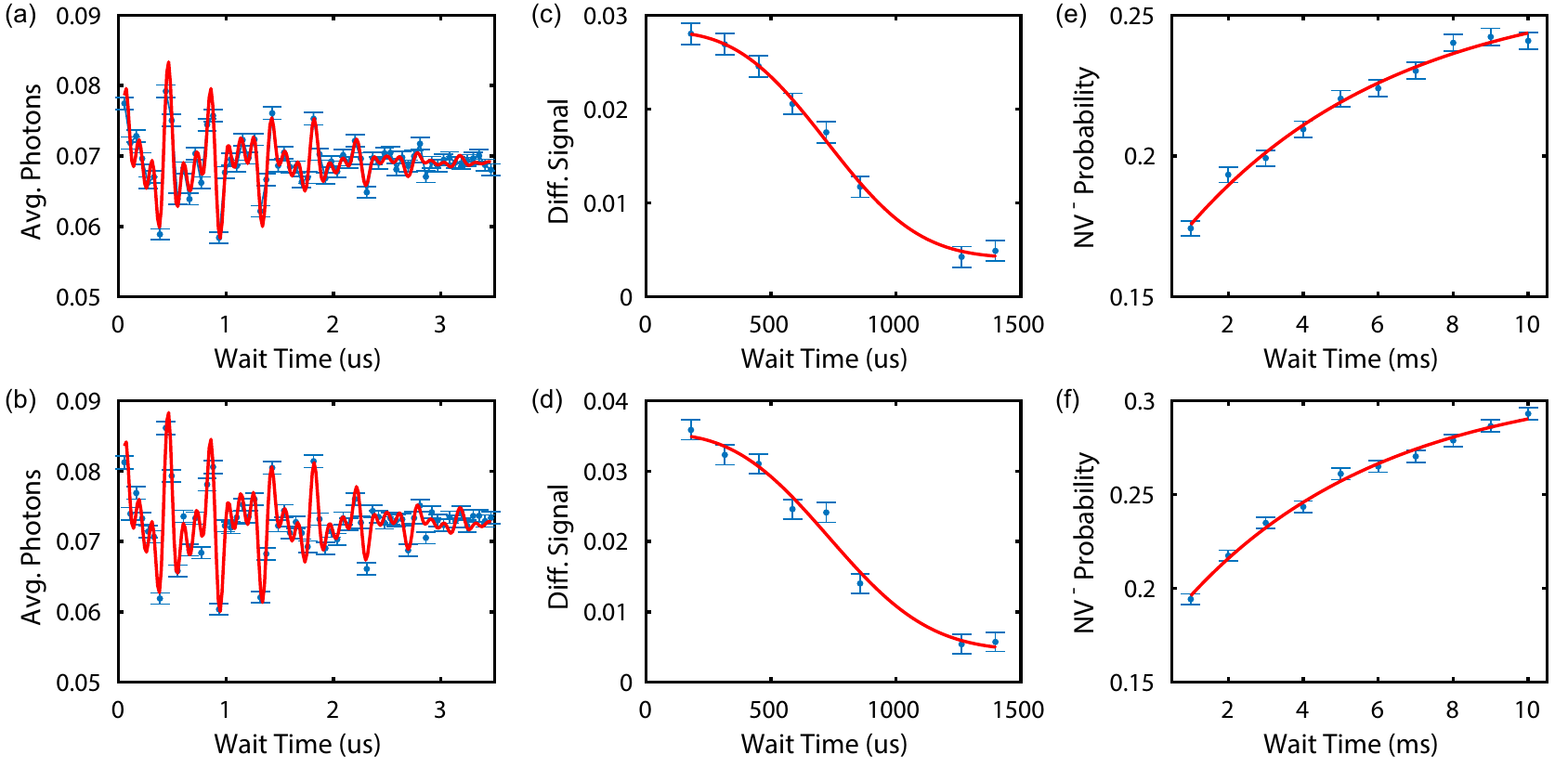}
\caption{Spin Property Comparison. Measurements of $T_2^*$ (a, b), $T_2$ (c, d), and $T_1$ (e, f) for the steady state and real-time initialization protocols, respectively.}
\label{sSFig:spin_properties}
\end{figure}

The ground state spin properties were measured for both the steady state and real-time initialization protocols.
First, we performed a Ramsey measurement on the NV center spin for both initialization techniques and measured the resulting spin population with traditional PL readout. 
The $\pi/2$ pulses were detuned from resonance by \SI{5}{\mega\hertz} to simplify the free evolution dynamics with the coupled $^{14}$N nuclear spin.
The data were fit to the model
\begin{equation}
    S = C + A*e^{-(\tau/T_2^*) ^2}\sum_{k = -1}^{1}\cos\left(2\pi(\delta - kA_{||}) + \phi\right),
\end{equation}
where $C$ is the dephased signal, $A$ is the amplitude of the signal, $\tau$ is the evolution time, $T_2^*$ is the inhomogenous dephasing timescale in \si{\micro\second}, $\delta$ is the detuning from resonance in \si{\mega\hertz}, $A_{||}$ is the parallel hyperfine coupling due to $^{14}$N in \si{\mega\hertz}, and $\phi$ is a phase offset in radians.
The data and fit results are depicted in SFig.~\ref{sSFig:spin_properties}(a, b) for the steady state and real-time initialization protocols, respectively.
We find a steady state $T_2^{*, SS}=1.92\pm0.08$\si{\micro\second} and a real-time $T_2^{*, RT}=2.24\pm0.10$\si{\micro\second}, which corresponds to a 16\% increase in spin dephasing time scale with the real-time initialization.

We then performed a Hahn-echo measurement for both of the initialization techniques. 
We initially identified the revivals due to the $^{13}$C spin bath and then measured the amplitudes of these revivals.
The final $\pi/2$ was applied around the $+X$ and $-X$ axes to allow for a differential signal, which can be seen in SFig.~\ref{sSFig:spin_properties}(c,d) for the steady state and real-time protocols, respectively.
The resulting data are fit to the model
\begin{equation}
    S = C + Ae^{-(\tau/T_2)^n},
\end{equation}
where $C$ is the offset, $A$ is the amplitude, $T_2$ is the echo coherence, and $n$ is a freely varying parameter for the stretched exponential \cite{Shields2015}.
The steady state initialization results in a coherence time of $T_2^{SS}=830\pm\SI{28}{\micro\second}$ and $n=2.98\pm0.42$.
The real time initialization results in a coherence time of $T_2^{RT}=852\pm\SI{81}{\micro\second}$ and $n=2.85\pm1.03$.

We also measured the spin relaxation time for both initialization techniques.
To do this, we first initialize into the $m_s=0$ projection and then record the SCC signal for various wait times.
The data is presented in SFigure~\ref{sSFig:spin_properties}(e, f) for the steady state and real-time initialization protocols, respectively.
The data in both cases is fit to a single exponential, which results in a spin relaxation time of $T_1^{SS}=5.6\pm1.2$\si{\milli\second} and $T_1^{RT}=5.3\pm0.9$\si{\milli\second} for the steady state and real-time protocols, respectively.
Thus, there is no measurable difference between the two techniques.

\section{Spin Readout Parameter Optimization}

\begin{figure}
    \centering
    \includegraphics{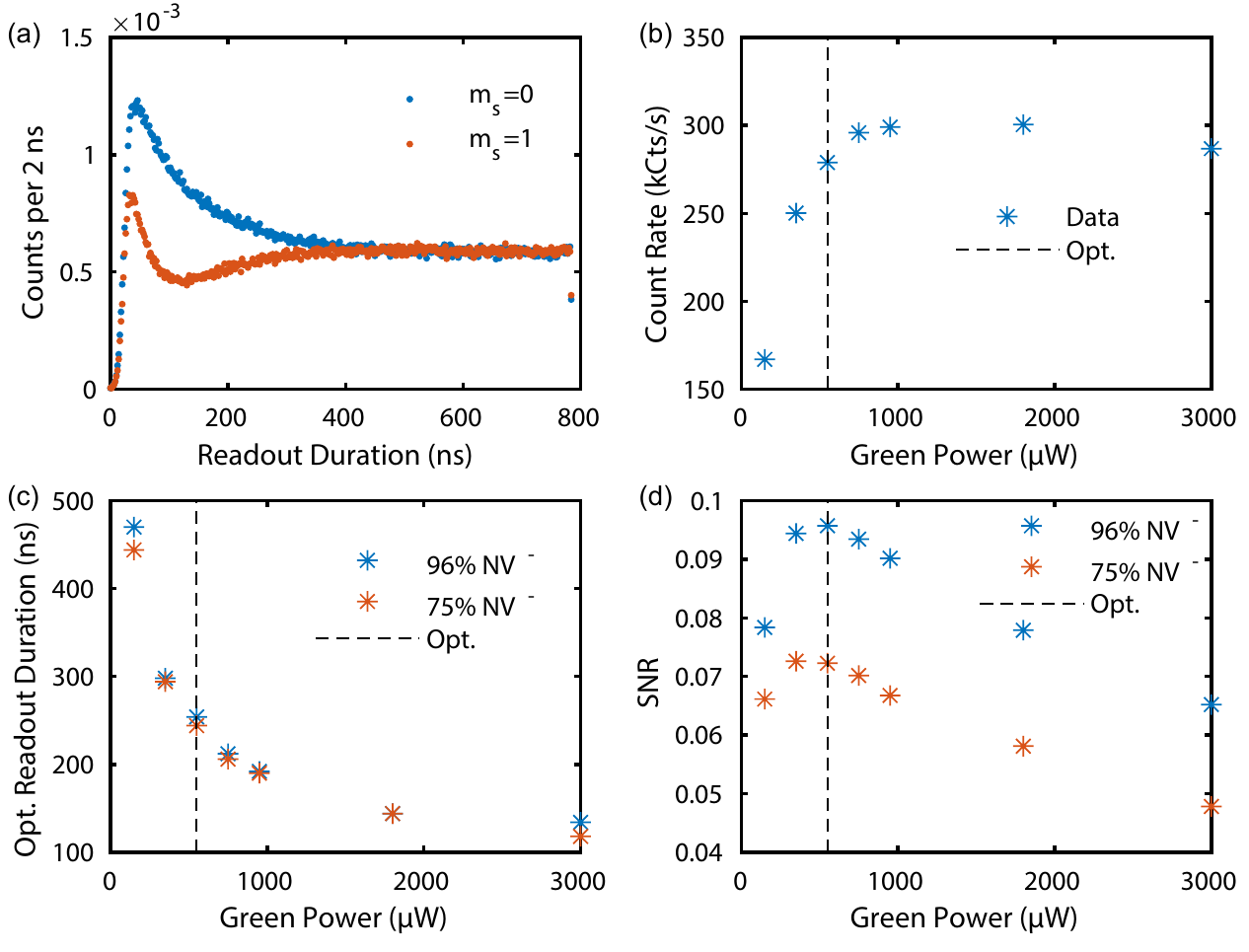}
    \caption{Traditional PL Readout Calibration (a) Example transient response due to the two spin states with  \SI{532}{\nano\meter} illumination. (b) The steady state count rate for different \SI{532}{\nano\meter} powers. (c,d) The optimum spin readout SNR and the corresponding power and readout duration as well as how they vary for different powers.}
    \label{SFig:pl_calibration}
\end{figure}

Both spin readout methods are calibrated to optimize their SNR to ensure that we are fairly comparing their performance.
The calibration of traditional PL readout is presented in SFigure~\ref{SFig:pl_calibration}.
We take time-tagged measurements of the transient fluorescence response to different \SI{532}{\nano\meter} illumination intensities [SFig.~\ref{SFig:pl_calibration}(d)] for both initialization techniques.
A few powers above and below the saturation power were chosen; note that we observed the quenching of fluorescence at higher powers [SFig.~\ref{SFig:pl_calibration}(b)].
From these time-traces, we can calculate the average number of detected photons for various integration windows and powers and find the global optimum.
We present how the optimum readout duration [SFig.~\ref{SFig:pl_calibration}(c)] and the optimum SNR [SFig.~\ref{SFig:pl_calibration}(d)] vary for different powers.
We find that both initialization techniques require the same readout duration and power, and find a strong quenching of the SNR for higher powers which coincides with the quenched saturation fluorescence, suggesting that a careful calibration of PL readout parameters is necessary when the NV is efficiently pumped.

\begin{figure}
    \centering
    \includegraphics{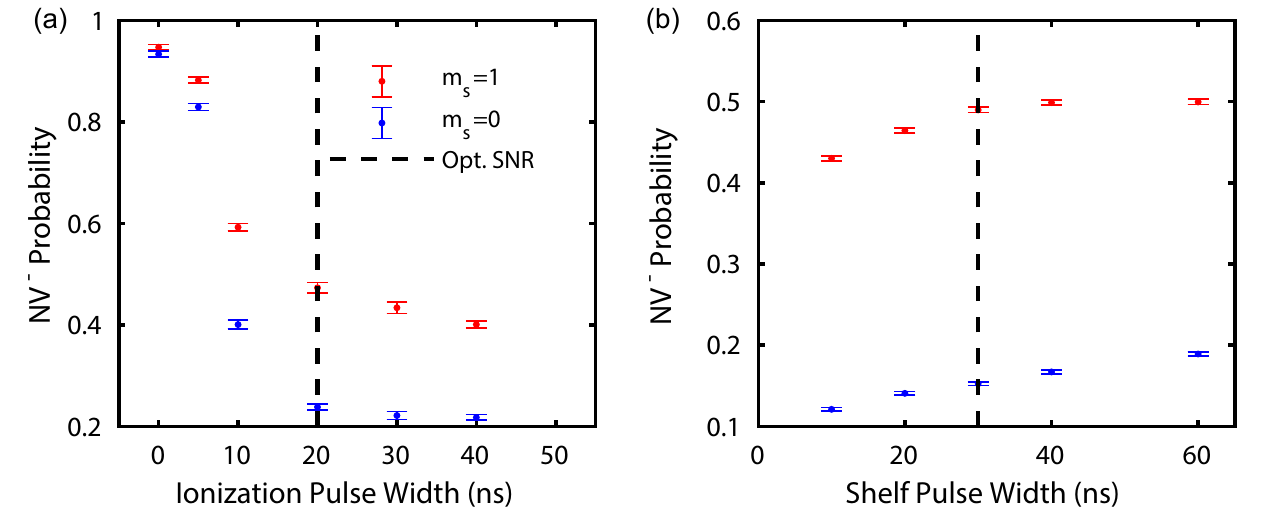}
    \caption{SCC Readout Calibration (a) NV$^-$ population for various ionization pulse widths (a) and shelf widths (b) for both spin states. The dashed black line signifies the optimal parameters for maximized single-shot SNR. }
    \label{SFig:scc_calibration}
\end{figure}

For spin-to-charge conversion (SCC), we follow a similar calibration scheme laid out in Supplemental Reference~\cite{Shields2015}, in which we sweep the durations of the orange shelf and red ionization pulse independently.
The approximate optimal shelf duration and power can be found by looking for the maximum instantaneous contrast between the $m_s=0$ and $m_s=\pm1$ states through a transient measurement [such as those in SFIG.~\ref{SFig:pl_calibration}(a)].
This initial calibration helps to reduce the large parameter space of the SCC readout properties.
For a given measurement sweeping over shelf or ionization duration, we simultaneously fit all of the charge readout distributions to the same underlying rates, but allow the NV$^-$ population for each spin and parameter to vary, thus reducing the covariance between NV$^-$ population and the ionization rate.
The results of the fitted NV$^-$ populations for these measurements are in SFig.~\ref{SFig:scc_calibration}.
We find an optimum ionization pulse settings of \SI{20}{\nano\second} and \SI{40}{\milli\watt} and a shelf settings of \SI{30}{\nano\second} and \SI{500}{\micro\watt}.
This produces a single-shot SNR=0.5 with an initial NV$^-$ fidelity of 98\%.

\section{Calibrating Spin SNR}
\begin{figure}
    \centering
    \includegraphics{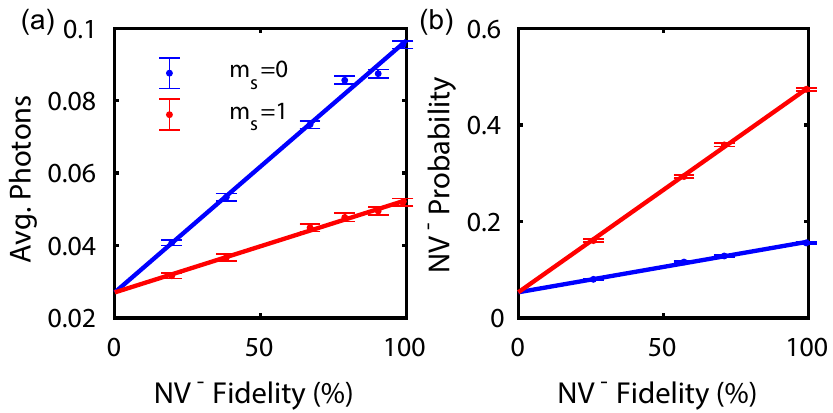}
    \caption{Fitting to spin observables (a) Average detected photons for both spin states and various initial NV$^-$ fidelities. (b) NV$^-$ probability following SCC for both spin states and various initial NV$^-$ fidelities. Solid lines are fits to Eqn. 6 in the main text where the error term is held constant between the two lines.}
    \label{SFig:spin_snr_fitting}
\end{figure}

The measured observables for different initialization fidelities and both readout techniques are presented in SFig.~\ref{SFig:spin_snr_fitting}.
For both readout techniques, the data are fit to Equation 6 in the main text, where we fix the error term to be the same for both initial spin states.
To be in line with the literature, we set $S=\alpha$ when our signal consists of average detected photons, and $S=\beta$ when our signal is the probability of detecting NV$^-$.
For PL readout [SFig.~\ref{SFig:spin_snr_fitting}(a)], the fitting yields $\braket{\Tilde{\alpha}_0'}=9.664(1)\times10^{-2}$ photons, 
$\braket{\Tilde{\alpha}_1'}=5.254(1)\times10^{-2}$ photons, and 
$\braket{\alpha_{\epsilon}}=2.703(3)\times10^{-6}$ photons.
For SCC readout [SFig.~\ref{SFig:spin_snr_fitting}(b)], the fitting yields $\braket{\Tilde{\beta}_0'}=0.1581(1)$ \%,
$\braket{\Tilde{\beta}_1'}=0.4778(2)$ \%, and 
$\braket{\beta_{\epsilon}}=0.0530(4)$ \%.

\section{Spin Polarization Estimation}

\begin{figure}
    \centering
    \includegraphics{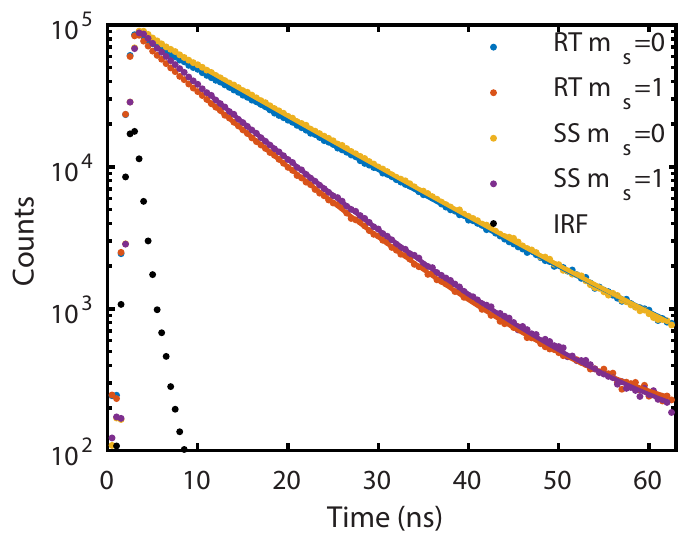}
    \caption{Lifetime Measurement of Spin Polarization. Transient response to a $<\SI{2.5}{\nano\second}$ red pulse for both the $m_s=0,1$ spin states and the real-time (RT) and steady state (SS) initialization protocols. The measured instrument response function (IRF) is also displayed. }
    \label{sfig:spin_polarization}
\end{figure}

The ground state spin polarization of the NV center can be estimated by observing the relative amplitudes of the bi-exponential decay of a pulsed lifetime measurement \cite{Fuchs2010, Robledo2011, Shields2015}.
By performing a fit to the lifetime response for both the optically polarized state, as well as after a calibrated $\pi$-pulse, one can estimate the initial $m_s=0$ population, assuming there is negligible radiative spin-mixing in the triplet manifold.
However, to accurately fit the lifetime response, one must take into account the other spin projection ($m_s=-1$ in our case), as well as pulse errors that manifest from the $^{14}$N hyperfine coupling.
We now write down the three spin projection populations before
\begin{align}
    \begin{pmatrix}
    p_{-1}^b \\
    p_{+1}^b \\
    p_0^b
    \end{pmatrix}
    =
    \begin{pmatrix}
    \frac{1-p_0}{2} \\
    \frac{1-p_0}{2} \\
     p_0
     \end{pmatrix},
\end{align}
and after the inversion pulse
\begin{align}
    \begin{pmatrix}
    p_{-1}^a \\
    p_{+1}^a \\
    p_0^a
    \end{pmatrix}
    =
    \begin{pmatrix}
    \frac{1-p_0}{2} \\
    \frac{1-p_0}{2}\times (1-F_\pi) + p_0 \times F_\pi   \\
    p_0\times (1-F_\pi) + \frac{1-p_0}{2}\times F_\pi
    \end{pmatrix},
\end{align}
where $F_\pi$ is the fidelity of the $\pi$-pulse, which for this measurement with a Rabi driving frequency of $\sim\SI{5}{\mega\hertz}$ is approximated to be 88\%.
The transient fluorescence response is then given by
\begin{align}
    f(t) = A_i\left(p_0^i e^{-\gamma_0t} + (p_{+1}^i + p_{-1}^i)e^{-\gamma_1t}\right) + C,
    \label{eqn:bi_exp}
\end{align}
where $A_i$ is an overall amplitude, $i=a,b$ represents whether this was after or before the inversion pulse, and $\gamma_i$ is the excited state decay rate for the magnitude of the spin projection $i$, and $C$ is a background term.

To perform the lifetime measurement, we initialize the NV center (either with the real-time or steady state protocol) and then measure the transient fluorescence response following a short excitation pulse with the red laser.
The red laser can be produce a pulse that is $\sim\SI{2}{\nano\second}$ in duration.
The instrument response function (IRF) for this can be seen in SFig.~\ref{sfig:spin_polarization}.
The data for both spin states and initialization protocols is presented in SFig.\ref{sfig:spin_polarization}.
We simulatenously fit Equation \ref{eqn:bi_exp}, convolved with the IRF, to the before and after inversion pulse data sets. 
This helps reduce covariance between the excited state decay rates and the relative amplitudes and has been utilized in the literature \cite{Fuchs2010}.
We determine the excited state lifetimes to be $\gamma_0^{-1}=12.50\pm\SI{0.02}{\nano\second}$, and $\gamma_1^{-1}=7.48\pm\SI{0.02}{\nano\second}$.
The spin polarization in $m_s=0$ ($p_0$) is found to be $91.5\pm0.7\%$ and $94.4\pm0.7\%$ for the steady state and real-time initialization techniques, respectively.

It should be noted that since our excitation pulse is of similar magnitude to the fastest decay rate ($\gamma_1^{-1} = \SI{7.5}{\nano\second}$), the spin polarization measured for the steady state technique may be slightly lower in actuality than measured, as the optical pumping process has begun during the non-instantaneous rise and fall time.
Nonetheless, we still measure a difference in the spin polarizations for the two techniques, which agrees with the independent observation of the SNR differences mentioned in the main text.

\section{Magnetic-field sensitivity}
We calculate the ac magnetic-field sensitivity for our NV using the expression

\begin{align}
    \eta_{\mathrm{AC}} = \frac{\pi\hbar}{2g\mu_B}\sqrt{\frac{T_2 + \tau_I + \tau_R}{(T_2)^2}}\sigma_R,
\end{align}
where $\tau_I$ is the initialization time, $\tau_R$ is the SCC readout duration, and $\sigma_R$ is the spin readout noise \cite{Hopper2018a}.
Our optimization routine maximizes the readout efficiency and provides us the initialization and readout durations as well as the single-shot SNR. 
The dependence of the spin readout noise on single-shot SNR, for thresholding, is given by \cite{Hopper2018a}

\begin{align}
    \sigma_R = \sqrt{1 + \frac{2}{\mathrm{SNR}^2}}.
    \label{eqn:spin_readout_noise}
\end{align}
The optimization routine for $\tau_O=\SI{800}{\micro\second}$ yields the following parameters: $\tau_I=\SI{43}{\micro\second}$, $P_\mathrm{Init.}=\SI{53}{\micro\watt}$, $\tau_R=\SI{127}{\micro\second}$, $P_\mathrm{Readout}=\SI{22}{\micro\watt}$ and $\mathrm{SNR}=0.4$. 
Using Eq.~\ref{eqn:spin_readout_noise}, we find $\sigma_R=3.67$.
This initialization duration corresponds to $F_{\mathrm{NV}^-}$ 98.7\%.
The SNR is below the maximum value reported value in the main text due to the shortened readout duration, and a subsequently reduced charge readout fidelity of 70\%.
This implies that, for this operation time, it is more advantageous to average more less accurate measurements.

\section{High Fidelity Initialization}
\begin{figure}
    \centering
    \includegraphics{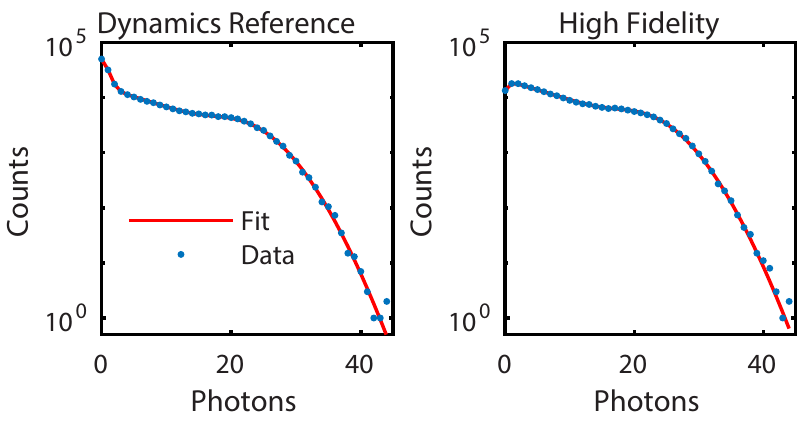}
    \caption{Measuring High Fidelity Initialization (a) The readout distribution resulting from a $73.3\pm0.2$\% NV$^-$ population and corresponding fit. (b) The readout distribution resulting from a $99.4\pm0.1$\% NV$^-$ population. The two curves are jointly fit to a model that fixes the charge dyamics but allows the initial NV$^-$ population to vary.}
    \label{sSFig:high_fid_init}
\end{figure}
High fidelity initialization can be achieved by increasing the threshold condition with similar probe power and durations that were presented in the main text.
To measure a high fidelity initialization attempt, we perform two interleaved measurements.
The first is a reference measurement where we measure the charge readout distribution due to a typical \SI{532}{\nano\meter} pump pulse, the second is the real-time initialization with charge probe settings of P$_\mathrm{probe}=$~\SI{6}{\micro\watt}, $\tau_\mathrm{probe}=$~\SI{9}{\micro\second}, and a threshold increased to $\nu=2$. 
We simultaneously fit to both the reference and high fidelity charge readout distributions where we fix the system dynamics between the two measurements ($\gamma_0, \gamma_-, \Gamma_\mathrm{Ion}$) and allow the NV$^-$ populations to vary. 
The elevated threshold has the effect of drastically increasing the average attempts but purifying the initial state.
We repeat the reference and real-time initialization measurement for 250,000 repeats, each; the results of this measurement are in SFigure~\ref{sSFig:high_fid_init}.
We determinee that the NV can be initialized into $F_{\mathrm{NV}^-}=99.4\pm0.1\%$ with an average time-to-initialize of \SI{7}{\milli\second}.
The maximum fidelity attainable with a threshold of 1 is $98.6\pm0.2\%$, which corresponds to an initialization error rate of 1.4\%, while a threshold of 2 results in an initialization error rate of 0.6\% which leads to a factor of 2.5 reduction in error rate.
In this regime, the initialization fidelity is limited by the signal-to-background ratio of the charge readout process.

\bibliography{si_bibliography}